 \documentclass[prb,twocolumn,showpacs,amsmath,amssymb,superscriptaddress]{revtex4-1}
\usepackage{graphicx}
\usepackage{bm}
\usepackage{bbm}

\usepackage{subfigure}
\usepackage{amssymb}

\newcommand{\beginsupplement}{%
        \setcounter{table}{0}
        \renewcommand{\thetable}{S\arabic{table}}%
        \setcounter{figure}{0}
        \renewcommand{\thefigure}{S\arabic{figure}}%
     }
     
\graphicspath{{DW_in_AFM_V2_graphics/}{DW_in_AFM_V2_tcache/}{DW_in_AFM_V2_gcache/}}
\DeclareGraphicsExtensions{.pdf,.svg,.eps,.ps,.png,.jpg,.jpeg}
\graphicspath {{DW_in_AFM_V2_graphics/}{DW_in_AFM_V2_tcache/}{DW_in_AFM_V2_gcache/}}
\DeclareGraphicsExtensions {.pdf,.svg,.eps,.ps,.png,.jpg,.jpeg}
\graphicspath {{DW_in_AFM_V1_graphics/}{DW_in_AFM_V1_tcache/}{DW_in_AFM_V1_gcache/}}
\DeclareGraphicsExtensions {.pdf,.svg,.eps,.ps,.png,.jpg,.jpeg}

\begin{document}

\title{Staggering antiferromagnetic domain wall velocity in a staggered spin-orbit field}

\author{O. Gomonay}
\affiliation{Institut f\"ur Physik, Johannes Gutenberg Universit\"at Mainz, D-55099 Mainz, Germany}
\affiliation{National Technical University of Ukraine ``KPI'', 03056, Kyiv,
	Ukraine}
\author{T. Jungwirth}
\affiliation{Institute of Physics ASCR, v.v.i., Cukrovarnicka 10, 162 53 Praha 6 Czech Republic}
\affiliation{School of Physics and Astronomy, University of Nottingham, Nottingham NG7 2RD, United Kingdom}
\author{J. Sinova}
\affiliation{Institut f\"ur Physik, Johannes Gutenberg Universit\"at Mainz, D-55099 Mainz, Germany}
\affiliation{Institute of Physics ASCR, v.v.i., Cukrovarnicka 10, 162 53 Praha 6 Czech Republic}

\begin{abstract}
	We demonstrate the possibility to drive an antiferromagnet domain-wall at high velocities by field-like N\'{e}el spin-orbit torques. 
	Such torques arise from  current-induced local fields that alternate 
	their orientation on each sub-lattice of the antiferromagnet and whose orientation depend primarily on the current direction, giving them their field-like character. 
	The domain-wall velocities that can be achieved 
	by this mechanism are two orders of magnitude greater than the ones 
	in ferromagnets. This arises  from the efficiency of the 
	staggered spin-orbit fields
	to couple to the order parameter and from the exchange-enhanced phenomena in antiferromagnetic texture dynamics, which 
	leads to a low domain-wall effective mass and the absence of a Walker break-down limit. 
	In addition, because of its nature, the staggered spin-orbit field 
	can lift the degeneracy between two  180$^\circ$ rotated   states
	in a collinear antiferromagnet and provides a  force that can move such walls and control the switching of the states. 
\end{abstract}

\maketitle



Antiferromagnets (AFs) are promising materials for spintronics because they show
fast magnetic dynamics, low  susceptibility to magnetic fields, and produce no stray fields. 
These advantages stem from the peculiarities of the AF ordering, 
which consists of alternating magnetic moments on individual atomic sites with  zero net magnetization, 
and whose orientation is described by the N\'{e}el vector. 
This also means  that an AF cannot be efficiently manipulated by external magnetic fields; 
a fact that has relegated AFs as primarily passive elements in todays technology. 
The emerging field of antiferromagnetic spintronics focuses on reversing this trend, making AFs active elements in spintronic based devices.\cite{Jungwirth2016,Editorial2016,Wadley2016,Marrows2016}

\begin{figure}[h]
	\centering
	\includegraphics[width=1.0\columnwidth]{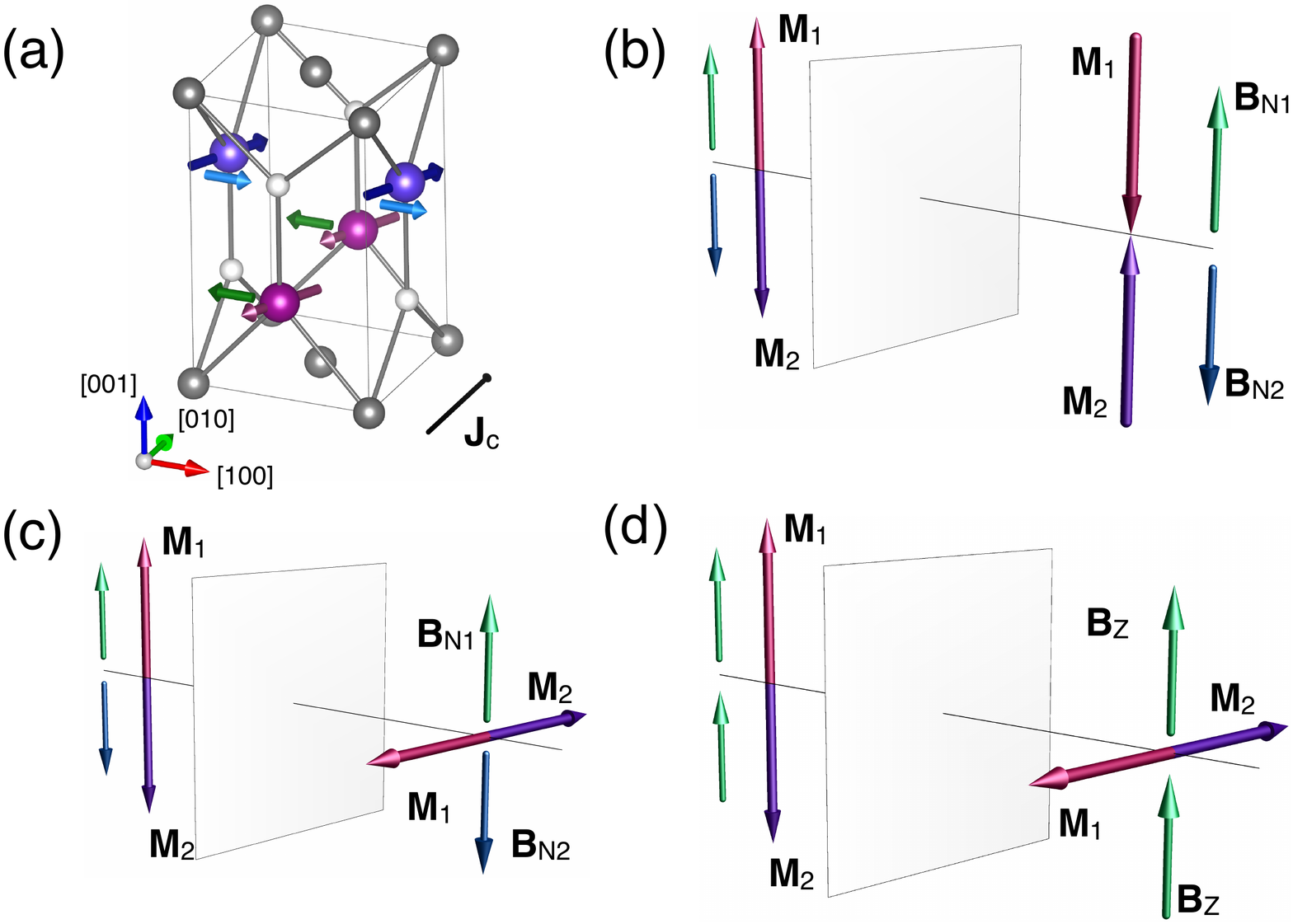}
	\caption[velocity]{
		(a) Crystal structure of antiferromagnetic  CuMnAs. The two spin-sublattices of the Mn atoms  are $\mathbf{M}_{1/2}$ (red and purple).
		The current-induced staggered NSOT
		field ($\mathbf{B}_\mathrm{N1/N2}$ - green and blue)
		has opposite sign at each Mn sublattice. 
		The non-magnetic As atoms (light grey) provide the locally broken  inversion symmetry at the Mn inversion partner sites which gives rise to the local current-induced fields via spin-orbit coupling. 
		$\mathbf{J}_{\rm c}$ represents the current. 
		(b) AFDW between 180$^\circ$ rotated antiferromagnetic domains in the presence of the staggered NSOT field. 
		The energy density has opposite sign in the left/right domains, producing a ponderomotive force. 
		(c-d) AFDW between 90$^\circ$ rotated antiferromagnetic  domains in (c) staggered NSOT field
		and (d) uniform Zeeman ($\mathbf{B}_{Z}$) field. 
	}
	\label{fig_1}
	\vspace{-0.5 cm}
\end{figure}

A new way to actively manipulate the N\'{e}el order parameter of the AF   is the 
recently proposed relativistic N\'{e}el spin-orbit torque (NSOT).\cite{Zelezny2014} 
This NSOT is the antiferromagnetic version of the inverse spin-galvanic (Edelstein) mechanism,\cite{Silov2004,Kato2004b,Ganichev2004b,Wunderlich2004,Wunderlich2005} which generates current-induced spin-orbit torques in ferromagnets (FMs).\cite{Chernyshov2009,Miron2010b}
It produces locally a non-equilibrium spin polarization in particular crystal structures that is proportional to the applied uniform current and alternates in sign
between the different magnetic sublattices. The local non-equilibrium spin polarization results in a staggered spin-orbit field that couples effectively to the N\'{e}el order parameter, as shown in Fig.~\ref{fig_1}(a).\cite{Zelezny2014,Wadley2016,note1} 

The NSOT arises in crystals whose magnetic atoms
have local environment with broken inversion symmetry and where the two magnetic
sublattices form inversion partners, such as in Mn$_2$Au and CuMnAs. Its first observation has been  recently reported in CuMnAs,\cite{Wadley2016}
with the measurements indicating that the NSOT switching involved a reconfiguration of a multiple-domain state of the AF. 
This motivates a study of current-induced AF dynamics beyond the coherent single-domain regime, 
in particular a study of the antiferromagnetic domain wall (AFDW) motion driven by the field-like NSOT. Both 90$^\circ$ and 180$^\circ$ AFDWs are experimentally relevant  since in thicker CuMnAs films an in-plane biaxial anisotropy dominates,\cite{Wadley2016} while thinner films (below $\sim$10~nm) are uniaxial.\cite{Wadley2015a}
In these crystals the staggered spin-orbit field is generated by an electrical current applied in the (001)-plane. The NSOT field is oriented in the plane in the direction perpendicular to the current and its amplitude from ab initio calculations is in the range of $\sim$~1-10~mT per 10$^7$~A/cm$^2$.

\begin{figure}[h]
	\centering
	\includegraphics[width=1.0\linewidth]{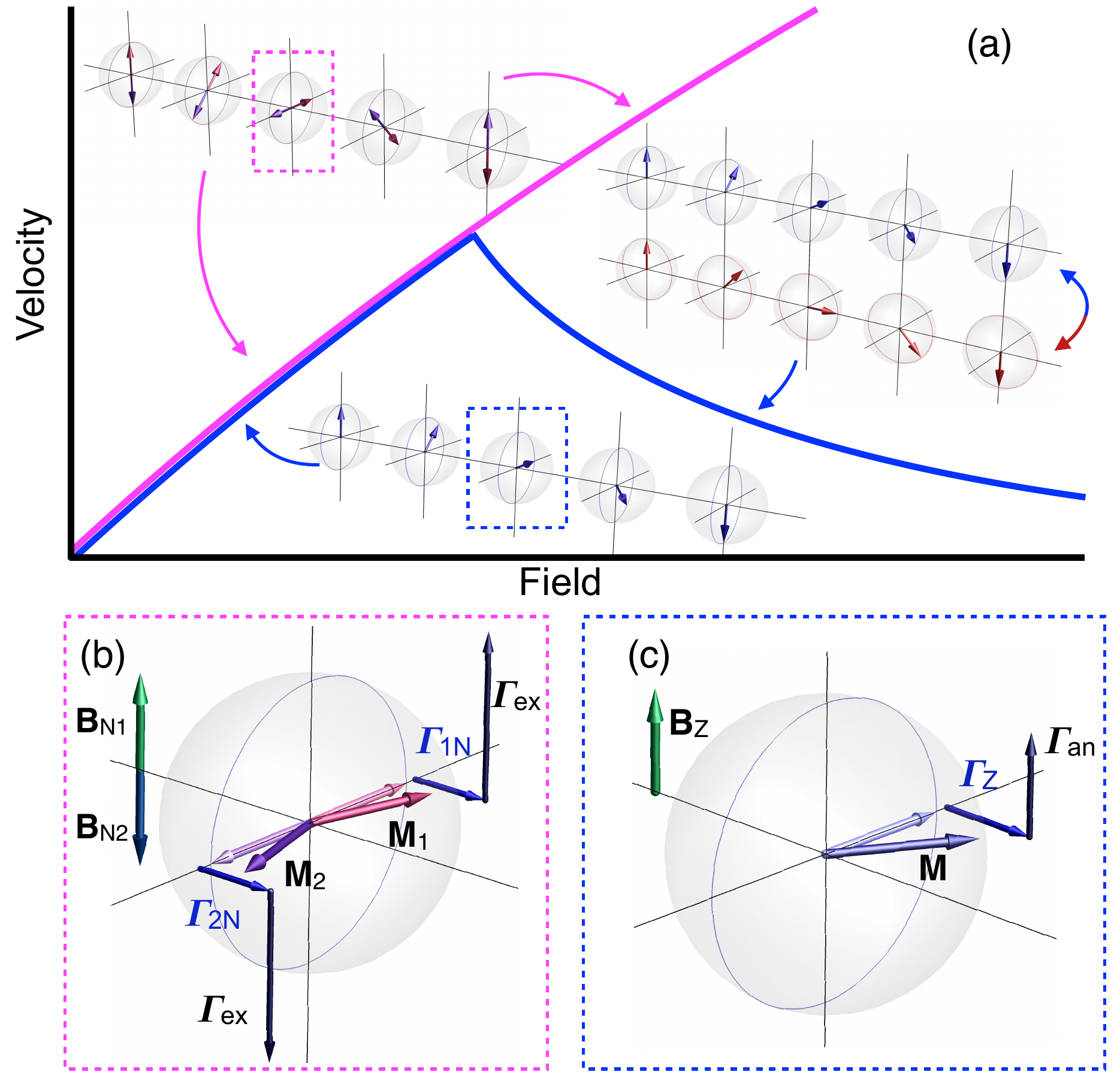}
	\caption[schema]{
		(a) Schematic of Bloch domain wall velocity vs. field for an AF (magenta) and a FM (blue). At low fields both FM and AF have
		the same velocity vs. field, but beyond the Walker-breakdown the FM slows down as the domain wall character oscillates between Bloch and N\'{e}el.
		(b-c) Illustration of torques exerted on the AFDW and FMDW center. (b) The N\'{e}el field generates a torque ($\mathbf{\Gamma}_{1N/2N}$) that cants $\mathbf{M}_{1/2}$ forward and the much larger internal torque  from the exchange interaction ($\mathbf{\Gamma}_{ex}$) rotates the sub-lattice magnetisations towards their respective easy axis directions which causes the AFDW motion. 
		The large ratio between exchange torque and external torque leads a to a very small deformation of the DW from its equilibrium (Bloch) configuration, i.e., to a very small AFDW mass.
		(c) In a FM the external driving field torque can reach similar magnitude as the 
		internal anisotropy torque ($\mathbf{\Gamma}_{an}$), which leads to larger deformation of the FMDW from its equilibrium (Bloch) configuration, i.e., to a much larger FMDW mass.
	}
	\label{fig_2}
	\vskip -0.5 cm
\end{figure}

In this Letter we present a theoretical study demonstrating  that AFDWs  can be  controlled electrically by the field-like NSOTs with high efficiency.
Moreover, the staggered NSOT field opens an unprecedented possibility to set into motion a 180$ ^\circ$ AFDW in a collinear AF. 
In addition, whereas the velocity of a ferromagnetic domain wall (FMDW) is limited by the Walker-breakdown,  in an AF  only the much higher magnon velocity sets the upper limit.\cite{Haldane1983a,Kosevich1990,Kim2014c} In a FM the Walker-breakdown arises when the  FMDW begins to oscillate between a Bloch and N\'{e}el type driven by a competition of the external field torque and the internal anisotropy torque, which can be of similar order, as shown in Fig.~\ref{fig_2} (a) and (c). In an AF on the other hand the internal torque driven by exchange is several orders of magnitude larger than any driving torque, which leads to a stiff AFDW, a very low effective mass of the AFDW, and no Walker-breakdown point, as shown in Fig.~\ref{fig_2} (a) and (b). 
At fields below the Walker-breakdown both the FM and AF have similar dependence on the driving field. 
In our calculations (assuming no extrinsic pinning) we show that the velocities are proportional to the N\'{e}el field and
estimate that they can reach values of $\sim 10-100$ km/s, orders of magnitude higher than in FMs. 
Also, by comparing the steady motion of a 90$^\circ$  AFDW  in the presence of a uniform Zeeman field and the staggered spin-orbit field, we 
show that the velocities induced by the latter are much
greater (for equal magnitudes of the two fields).

Experimentally it has been shown that the AFDW can be dragged by an STM-tip that generates a spin-polarized current, with the velocity of the AFDW equal to the
velocity of the tip.\cite{Wieser2011} 
There have also been other mechanisms proposed for the induced motion of the AFDW. 
It can be pushed by circularly polarized magnons.\cite{Kim2014c} 
In systems formed by antiferromagnetically coupled
thin-film FMs  (synthetic AFs), where the antiferromagnetic  coupling is not as strong as in our case of bulk AF, 
an antiferromagnetic texture can be moved by the electric current due to
dissipative (antidamping-like) spin torques.\cite{Hals2011,Saarikoski2014,Yang2015a}
Another proposed method to manipulate an AFDW is
with the gradient of external magnetic field.\cite{Tveten2015a}
All of these proposed methods for AFDW manipulation cannot reach the high velocities and efficiencies afforded by the field-like NSOTs.

We consider a compensated collinear AF described by the sublattice magnetization
vectors $\mathbf{M}_{1}$ and $\mathbf{M}_{2}$ ($\left \vert \mathbf{M}_{1}\right \vert =\left \vert \mathbf{M}_{2}\right \vert =M_{s}/2$),
in which the electrical current generates the non-equilibrium  staggered
spin-orbit field described by  vectors $\mathbf{B}_\mathrm{N1}$ and $\mathbf{B}_\mathrm{N2}$ acting on $\mathbf{M}_{1}$ and $\mathbf{M}_{2}$, respectively,
as shown in Fig.~\ref{fig_1} (a). For convenience we measure $%
\mathbf{B}_\mathrm{N1/N2}$ in units of the magnetic field. The conversion to current
density  
can be calculated by microscopic techniques as, e.g.,  in Ref. \onlinecite{Zelezny2014}. 

Due to the direct coupling between the atomic spins and the local fields,
the current-induced contribution to the magnetic energy density of an AF takes a form $w = -\mathbf{M}_{1} \cdot\mathbf{B}_\mathrm{N1} -\mathbf{M}_{2} \cdot\mathbf{B}%
_\mathrm{N2}=-\mathbf{L} \cdot \mathbf{B}_\mathrm{N e e l}$, where $\mathbf{L} \equiv\mathbf{M}_{1} -\mathbf{M}_{2}$ is the
N\'{e}el order parameter vector, and $\mathbf{B}_\mathrm{N e e l}\equiv(\mathbf{B}_\mathrm{N1} -\mathbf{B}_\mathrm{N2})/2$. The sign and orientation of $\mathbf{B}_\mathrm{N e e l}$ is dermined by the sign and orientation of the applied electrical current.
In the presence of an external uniform magnetic field or in a general case when $(\mathbf{B}_\mathrm{N1} +\mathbf{B}_\mathrm{N2})/2\neq 0$, there is an additional uniform Zeeman field contribution, $\mathbf{B}_\mathrm{Zee}$,  which in the magnetic energy density expression 
couples to the uncompensated magnetization of the AF, $\mathbf{M}_\mathrm{A F} =\mathbf{M}_{1} +\mathbf{M}_{2}$. 
The externally applied Zeeman and N\'{e}el fields act
on an AF in different ways. The N\'{e}el field can change only the equilibrium orientation of
the AF vector $\mathbf{L}$. 
On the other hand, the Zeeman field produces a
small magnetization $\mathbf{M}_{A F} =\mathbf{L} \times \mathbf{B}_\mathrm{Zee} \times \mathbf{L}/(M_{s} H_\mathrm{ex})$, where $H_\mathrm{ex}$
stands for exchange field that keeps magnetic sublattices antiparallel. 

The final expression for the magnetic energy density  can then be written as: 
\begin{equation}
w = -\frac{1}{2 M_{s} H_\mathrm{ex}} \left (\mathbf{L}
\times \mathbf{B}_\mathrm{Z e e}\right )^{2} -\mathbf{L} \cdot\mathbf{B}_\mathrm{%
	Neel} .  \label{eq_field_energy}
\end{equation}
It follows from Eq.~(\ref{eq_field_energy}) that the effect produced in AFs by
the Zeeman field is i) quadratic in $\mathbf{B}_%
\mathrm{Z e e}$ and ii) weakened due to the strong exchange interaction. In contrast, 
the effect of the N\'{e}el field is linear in $\mathbf{B}_%
\mathrm{N e e l}$ and not diminished by the strong exchange interaction. Hence 
its effect will be much stronger than the effect of the Zeeman field. 
It is also important to note that the N\'{e}el field can remove the degeneracy of states with opposite direction of 
$\mathbf{L}$, while other physical fields
can distinguish only between states with different
orientation of $\mathbf{L}$. This directly implies  that
the N\'{e}el field can produce an effective  force per area ~$2 \mathbf{L}\cdot \mathbf{B}_\mathrm{N e e l}$
that will set into motion the AFDW between 180$^\circ$ rotated domains that is independent of the microscopic structure of the AFDW. The sign of $\mathbf{B}_\mathrm{N e e l}$ (of the applied electrical current) determines the direction of the AFDW motion.

To study this problem in more detail we consider  an example of a one-dimensional 
texture in a uniaxial AF and  in the presence of a dc N\'{e}el field parallel to the AF
easy axis (see Fig.~\ref{fig_1} (b)). 
Such AF has two  states that are magnetically equivalent at zero fields with $\mathbf{L}_{1} = -\mathbf{L}_{2}$
parallel to the easy axis. Both states have the same Zeeman energy, since $%
\left (\mathbf{L}_{1} \times \mathbf{B}_\mathrm{Z e e}\right )^{2} =\left (\mathbf{L}_{2}
\times \mathbf{B}_\mathrm{Z e e}\right )^{2}$, and therefore the Zeeman field can be neglected. The dynamics
of an AF texture is described by  phenomenological equations  for the AF vector  (see, e.g. Refs. \onlinecite%
{Baryakhtar1979, Andreev1980, Baryakhtar1980}). In our
case these equations are reduced to the following equation for the angle $\theta (x ,t)$
between $\mathbf{L}$ and the easy axis: 
\begin{eqnarray}
c^{2} \frac{\partial^2 \theta}{\partial x^2 } -\ddot{\theta } -\gamma ^{2} H_\mathrm{ex} H_%
\mathrm{a n} \sin \theta \cos \theta \nonumber \\
=\alpha _{G} \gamma H_\mathrm{ex} \dot{%
	\theta } +\gamma^2 H_\mathrm{ex} B_\mathrm{N e e l} \sin \theta ,
\label{eq_equation_theta_Neel_field}
\end{eqnarray}
where $\gamma $ is gyromagnetic ratio, $H_\mathrm{a n}$ is the magnetic
anisotropy field, $c$ is the magnon velocity, and $\alpha _{G}$ is the Gilbert
damping parameter.

Equation (\ref{eq_equation_theta_Neel_field}) has a solution which describes
a moving AFDW separating domains with $\theta _{1} =0$ and $\theta_{2}
=\pi $. The velocity of steady motion, 
\begin{equation}
v_\mathrm{s t e a dy}^\mathrm{AF} =\frac{2B_\mathrm{N e e l} c}{\sqrt{\alpha
		_{G}^{2} H_\mathrm{a n} H_\mathrm{ex} +4B_\mathrm{N e e l}^{2}}},
\label{eq_velocity_uniaxial_AFM_1}
\end{equation}
is obtained from the balance between the  force produced by the N\'{e}el
field and the internal (Gilbert) damping. In contrast to the ferromagnetic case, the velocity is only limited by
the magnon velocity $c=\gamma \sqrt{A H_\mathrm{ex}/M_s}$ (as mentioned in
Refs. \onlinecite{Haldane1983a,Kosevich1990,Kim2014c}), where $A$ is the
exchange stiffness.

It is instructive to compare this result with the steady motion of the FMDW, separating 180$^\circ$  rotated domains
in a uniaxial FM, induced by a Zeeman magnetic field or, equivalently,
by the field-like component of a current-induced spin torque  (see e.g. Ref.~\onlinecite{Seo2012} for details).
Such a FMDW cannot move while keeping its form, even in the
presence of a Zeeman field parallel to the easy axis, since a parallel shift is
connected with a variation of the total magnetization.\cite{Kosevich1990} In contrast,
the magnetization of an AF in the presence of the N\'{e}el field has a pure dynamic
origin. Hence, the parallel shift of an AFDW does not affect the total magnetization of the texture.

Steady motion of the FMDW in a uniaxial FM is often combined with the
rotation of the magnetization around the easy axis with a constant angular 
velocity $\omega ={\gamma B_\mathrm{Zee}}/(1 +\alpha _{G}^{2})$.  In this
case the velocity of the steady motion is proportional to the damping
coefficient:  
\begin{equation}
v =\frac{\gamma \alpha _{G} x_\mathrm{D W} B_\mathrm{Zee}}{1 +\alpha _{G}^{2}%
},  \label{eq_velocity_FM_uniaxial}
\end{equation}
where $x_\mathrm{D W}=\sqrt{A/H_\mathrm{a n} M_s}$ is the wall width.
In the more realistic case considered by  Walker,\cite{Schryer1974} the magnetic
anisotropy function includes demagnetization energy. In this case,
magnetization in the moving FMDW makes a constant angle $\sin 2 \varphi _{0} ={%
	H_{\rm Zee}}/{H_{c}}$ with the FMDW plane, where the critical field $H_{c} \approx \alpha_{G} H_\mathrm{dip-an}$ sets the Walker limit for the FMDW velocity (we assume that the dipolar anisotropy field $H_\mathrm{dip-an}$ is comparable to the anisotropy fields):  
\begin{equation}
v_\mathrm{steady}^\mathrm{FM} = \frac{\gamma B_\mathrm{Zee} x_\mathrm{D W}}{%
	\alpha _{G} \sqrt{1 +\frac{H_\mathrm{dip-an}}{H_\mathrm{a n}} -\left (\frac{H_\mathrm{dip-an}}{H_%
			\mathrm{a n}}\right ) \sqrt{1 -\frac{B_\mathrm{Zee}^{2}}{H^2_{c}}}}}.
\label{eq_velocity_FM_biaxial}
\end{equation}

The mobilities of 
AFDWs and of FMDWs below the Walker limit
could be of the same order for systems with similar values of the wall
width and Gilbert damping.
This can be seen by comparing Eq.~(\ref{eq_velocity_uniaxial_AFM_1})  and Eq.~(\ref%
{eq_velocity_FM_biaxial}):  
\begin{eqnarray}  \label{eq-mobility_DW}
\mu^\mathrm{FM}&\equiv&\frac{dv_\mathrm{steady}^\mathrm{FM} }{dB_\mathrm{Zee}}=%
\frac{\gamma x_\mathrm{D W}}{\alpha_G},
\\
\mu^\mathrm{AF}&\equiv&\frac{dv_\mathrm{%
		steady}^\mathrm{AF} }{dB_\mathrm{Neel}}=\frac{c}{\alpha_G\sqrt{H_\mathrm{a n}
		H_\mathrm{ex}}}\approx\frac{\gamma x_\mathrm{D W}}{\alpha_G}.\label{eq-mobility_AFDW}
\end{eqnarray}
However, the limiting velocity of the AFDW coincides  with the magnon
velocity, $v^\mathrm{AF}_\mathrm{l i m} =\gamma \sqrt{H_\mathrm{ex} A/M_s}$,
which due to strong exchange enhancement, is much larger than the typical magnon
velocity in a FM. On the other hand, in a FM the limiting (Walker) velocity, $v^\mathrm{FM}_\mathrm{l i m}\approx \gamma \sqrt{H_\mathrm{dip-an}A/M_s}$,
where we have assumed $H_\mathrm{dip-an}$ is of the order of $H_\mathrm{an}$.
Hence, $v^\mathrm{FM}_\mathrm{l i m}$ is  much smaller than $v^\mathrm{AF}_%
\mathrm{l i m}$. For example, typical values of $v^\mathrm{AF}_\mathrm{l i m}%
=c$ vary from 36 km/s in dielectric NiO, \cite{Hutchings1972} 40-50 km/s
in metallic $\gamma-$Mn$_{1-x}$Cu$_x$ alloys \cite%
{Cywinski1980,Wiltshire1983, Wiltshire1983a}, and up to 90 km/s for an AF KFeS$_2$
with extremely large magnon frequency (10 THz).\cite{Welz1992} For
comparison, the highest FMDW velocities 
range from 100 m/s \cite{Beach2005} to 400 m/s \cite{Miron2011} and a
velocity up to 750 m/s was recently achieved in a synthetic AF.\cite{Yang2015a}



In order to illustrate the efficiency of the NSOT we compare next the effects of the N\'{e}el and Zeeman fields on AFDWs. To do so it is better to reduce the complexity of Eq.~(\ref{eq_equation_theta_Neel_field}), which describes
fully the dynamics of the AF texture in all space, to one where the AFDW can be treated as a point particle with an effective mass. This will be the case if, e.g., the AFDW thickness is much smaller than the sample dimensions. In such cases the motion can be described with a reduced number of dynamical variables. The most natural way to introduce these variables is through the integral of motion for the moving AFDW in the absence of external forces (i.e., in a homogeneous isotropic space). Then the integral of motion related with translation invariance is the momentum of the AFDW (for exact expression see Eq. (6) Supplementary material).
Variation of momentum  is due to the presence of external forces which break translational invariance of space (ponderomotive forces) and time inversion (dissipative and gyrotropic forces).

Following this derivation, the equation for the AFDW momentum is given by $P_{x}\propto -\int (\partial\theta/\partial x) \dot{\theta}dx$, \cite{Haldane1983a}  
instead of the explicit Eq.~(\ref{eq_equation_theta_Neel_field}) for the
AFDW profile. Doing this one obtains that for a steady moving texture, $P_{x}\propto v/\sqrt{1-v^{2}/c^{2}}$, i.e. dynamics is Lorentz invariant.\cite{Haldane1983a,Kosevich1990}
The corresponding equation of motion takes a form 
\begin{equation}
\frac{dP_{x}}{dt}=-\alpha _{G}\gamma H_{\mathrm{ex}}P_{x}+F_{x},
\label{eq_momentum_general}
\end{equation}
where $F_{x}$ is the effective force which we specify below for each
case. The detailed derivation of Eq.~(\ref{eq_momentum_general}) is
given in the Supplementary material, as well as  explicit expressions for external forces and the AFDW mass, $M_\mathrm{AFDW}\propto1/H_{\rm ex}$. 
From Eq.~(\ref{eq_momentum_general}) the relaxation time of the AFDW is given by $\tau_{\rm{AF}}=1/\alpha_G\gamma H_{\rm ex}$. While both the mass and relaxation time are strongly suppressed in the AF due to exchange, the ratio $M_\mathrm{AFDW}/\tau_\mathrm{AF}=\alpha_GM_sS/\gamma x_\mathrm{DW}$ is independent of $H_{\rm ex}$ and is the same as in the FMDW (see Supplementary material). This implies the  AFDW and FMDW mobilities below the Walker breakdown are comparable, as shown earlier in Eqs.~(\ref{eq-mobility_DW}) and (\ref{eq-mobility_AFDW}).

We can understand Eq.~(\ref{eq_momentum_general}) intuitively from its Lorentz invariant character. 
Because a shift of the domain wall needs some energy for
the reorientation of magnetic moments, its inertia  is proportional
to the wall width. 
However, due to the relativistic character of the AF dynamics, the width of the moving AFDW depends upon its velocity (it shrinks proportional to a factor $\sqrt{1-v^{2}/c^{2}}$).   Hence, 
Eq.~(\ref{eq_momentum_general}) can be treated as the equation of motion for a particle moving in a Lorentz-invariant system under the action of viscous damping (the first term in r.h.s.) and an effective force (the second term in r.h.s.).

\begin{figure}[h]
	\centering
	\includegraphics[width=1.0\columnwidth]{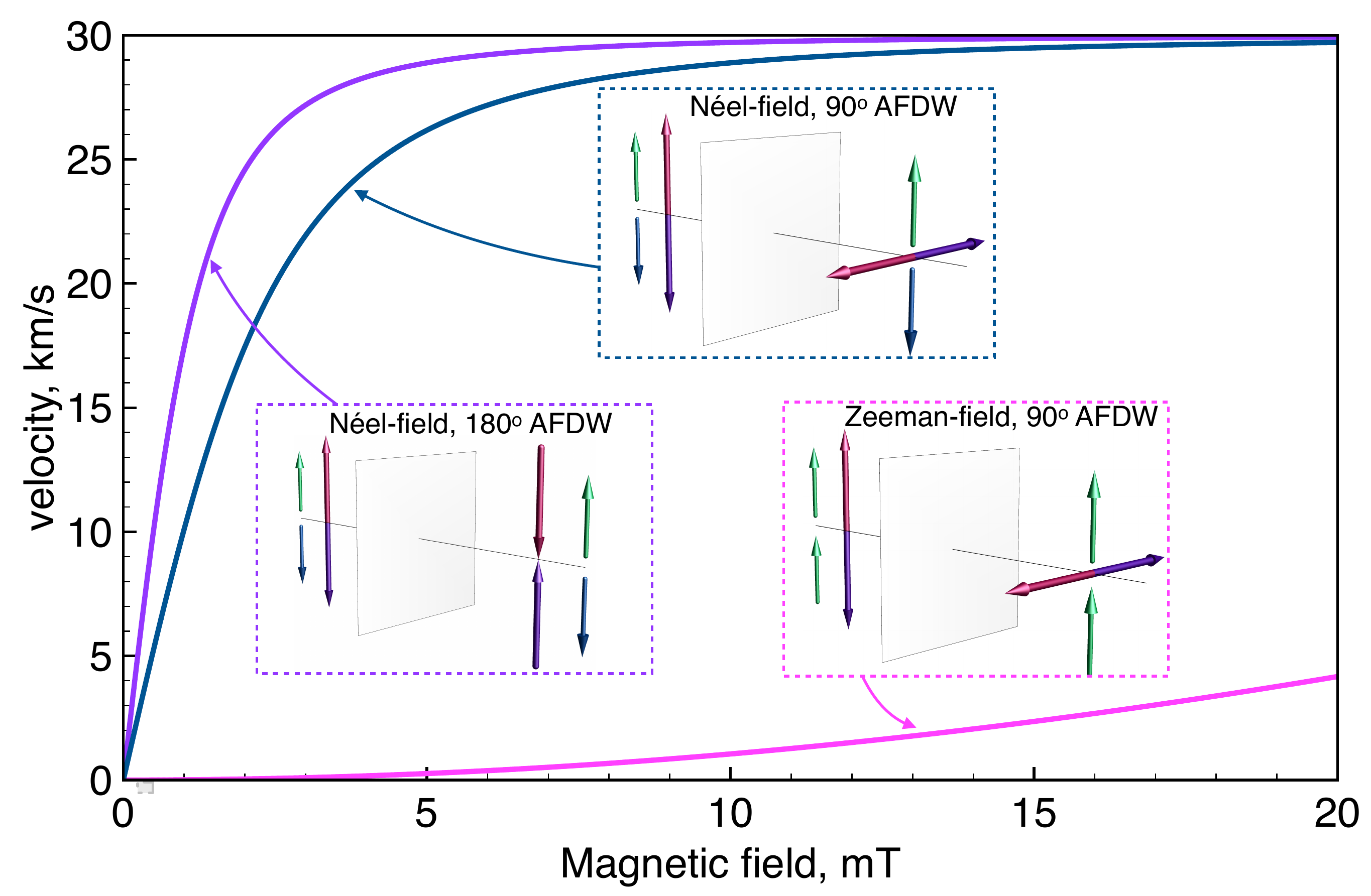}
	\caption[fig3]{Velocity of a $180^\circ$ (violet line) and a $90^\circ$  (green-blue line) AFDW vs effective N\'{e}el field  and a $90^\circ$ AFDW(magenta line)  vs Zeeman field  calculated for  Mn$_2$Au  (we set AF magnon velocity $c=30$ km/s).}
	\label{fig_3}
\end{figure}

To compare the effects of static N\'{e}el and Zeeman fields  we
consider the dynamics of the 90$^{\circ }$ AFDW. Both fields remove
the degeneracy of the states $\mathbf{L}_{1}\perp \mathbf{L}_{2}$ and thus could
produce the effective force per area, 
$F_x=w(\mathbf{L}_{1})-w(\mathbf{L}_{2})$ (see also Fig.~\ref{fig_1}(c),(d)). The possible ranges of the fields are limited
by the critical values (monodomainization field) at which one of the equilibrium
states disappears: by the spin-flop field, $H_{\mathrm{sf}}=\sqrt{H_{\mathrm{an}}H_{\mathrm{ex}}}$, in the case of  Zeeman field and by $H_{\rm an}$ in the case
of  N\'{e}el field. If both fields are applied parallel to one of the easy
axis (Fig.~\ref{fig_1}(c),(d)), they can compete with or add  to each other, depending on the sign of the N\'{e}el field (sign of the applied current), and the velocity of steady motion is 
\begin{equation}
v_\mathrm{steady}^\mathrm{AF}=c\frac{B_\mathrm{Neel}-B_\mathrm{Zee}^{2}/(2H_{\mathrm{ex}})}{\sqrt{\alpha _{G}^{2}H_{\mathrm{an}}H_{%
			\mathrm{ex}}+[B_\mathrm{Neel}-B_\mathrm{Zee}^{2}/(2H_{\mathrm{ex}})]^{2}}},
\label{eq_velocity of steady motion_parallel}
\end{equation}%
as can be obtained from Eq. (\ref{eq_momentum_general}). Fig.~\ref{fig_3} shows the field dependence of the velocity calculated according to Eq.~(\ref{eq_velocity_uniaxial_AFM_1}) (180$^\circ$ AFDW) and 
Eq.~(\ref{eq_velocity of steady motion_parallel}) (90$^\circ$ AFDW) for Mn$_{2}$Au taking the exchange field $H_\mathrm{ex}$=1307 T, 
\cite{Wu2012} in-plane magnetic anisotropy $H_\mathrm{a n}$
=0.03 T, \cite{Shick2010} and setting $%
\alpha_G=10^{-3}$. 

\begin{figure}[h]
	\centering
	\includegraphics[width=1.0\columnwidth]{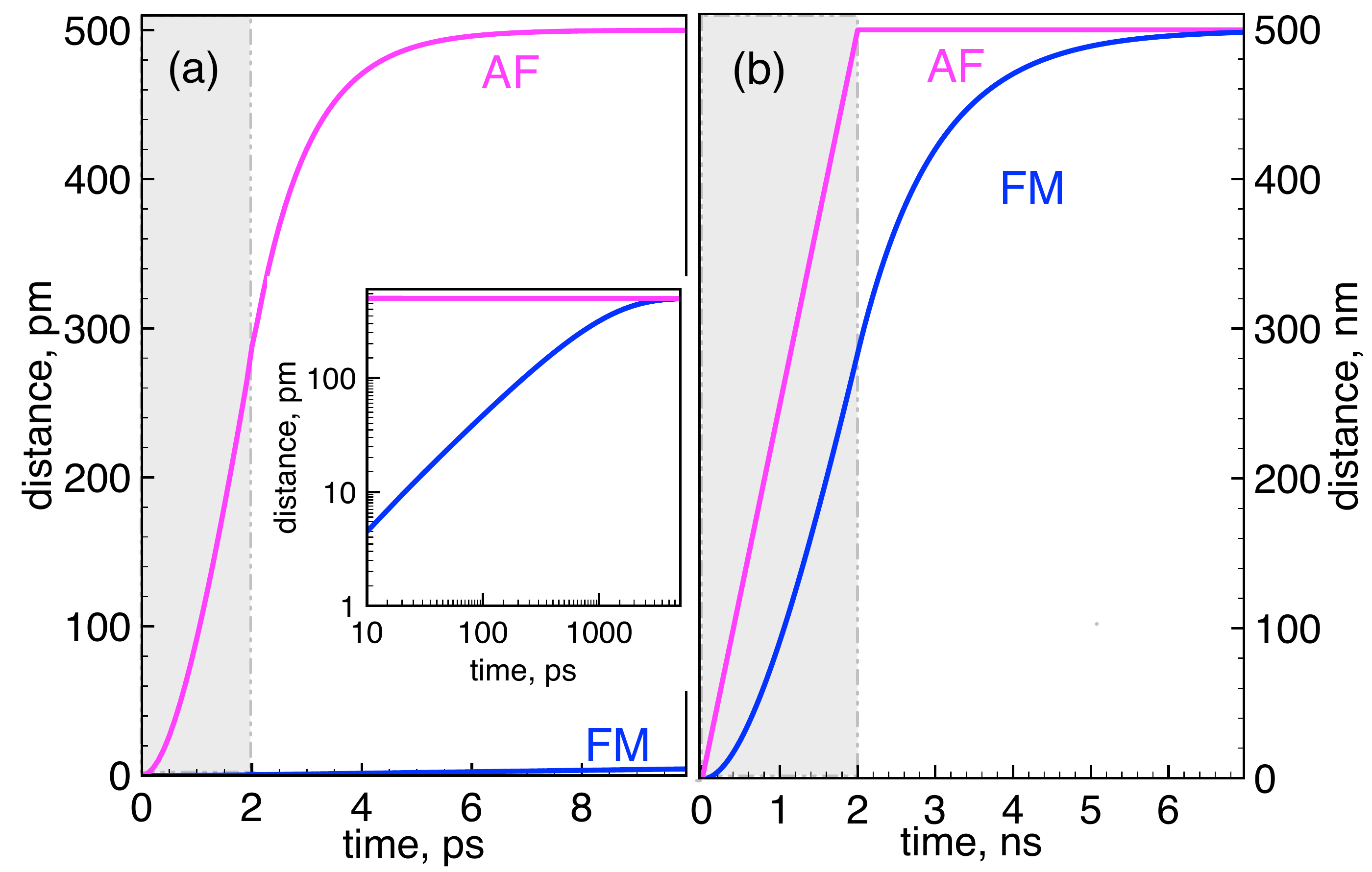}
	\caption[fig4]{Shift of (a) an AFDW (magenta lines)  and (b) a FMDW (blue lines) under the action of a field pulse (duration of pulse shown as grey region). Field values in both cases correspond to the steady velocity of 250 m/s in the dc field. Relaxation times are $\tau_{\mathrm{AF}}=1$~ps and $\tau_{\mathrm{FM}}=1$~ns for AF and FM respectively. In (a) the pulse duration $\tau=2$~ps is much smaller than the relaxation time of FM ($\tau_{\mathrm{FM}}$) but larger than $\tau_{\mathrm{AF}}$. Inset shows the time dependence at large time scales. In (b) the pulse duration $\tau=2$~ns is  larger than $\tau_{\mathrm{FM}}\gg\tau_{\mathrm{AF}}$. 
	}
	\label{fig_4}
\end{figure}

The three orders of magnitude  difference in the effective force occurs due to the exchange reduction of the Zeeman-field effects in an AF. As a result, the contribution of the Zeeman field to the domain wall velocity (magenta line in Fig.~\ref{fig_3}) is vanishingly small compared to  the contribution of the N\'{e}el  field (green-blue line). We note that experimentally the effectiveness of the N\'{e}el field compared to the Zeeman field has been observed for current-induced 
reconfiguration of 90$^o$ rotated domains in AF CuMnAs.\cite{Wadley2016}.  According to  microscopic calculations of the current induced N\'eel field in CuMnAs,\cite{Wadley2016}  the AF states were  switched by a current which corresponds to $B_\mathrm{Neel}\sim $1 mT, while the Zeeman field up to 12 T was not sufficient for such switching.

It is also worth noting that the maximum domain wall velocity observed up to now in the synthetic AFs  was at current densities $3\times10^8$A/cm$^2$.\cite{Yang2015a} According to our calculations, the same AFDW velocity  in bulk AF Mn$_2$Au corresponds to a staggered spin-orbit field of 0.07~mT which, according to Ref.~\onlinecite{Wadley2016}, corresponds to a current density of 3.5$\times 10^5$ A/cm$^2$.   

Another experimentally relevant calculation that highlights the efficiency of the NSOT field is  the domain wall displacement driven by short pulses. We present the results in Fig.~\ref{fig_4}, where a single rectangular
pulse is delivered with a field magnitude corresponding to a stead-state domain wall velocity of 250 m/s (the maximum FMDW velocity reached in experiment\cite{Jiang2010} ). For the AFDW we consider the N\'eel field, for the FMDW the Zeeman field and the relaxation times for the  AF and FM are $\tau_{\mathrm{AF}}=1$~ps and $\tau_{\mathrm{FM}}=1$~ns, respectively. 
Although the ultimate displacement of both the FMDW and the AFDW is the same, the AFDW attains it much faster due to its low mass and the resulting weak AFDW inertia. The favorable characteristics of the AFDW dynamics compared to the FMDW is more pronounced at short pulses ( $\ll\tau_{\mathrm{FM}}$), as shown in  Fig.~\ref{fig_4}. 

In summary, we have demonstrated the efficiency of the AFDW motion by the NSOT field. The limiting velocity of  the AFDW motion induced by the N\'{e}el field is several orders of magnitude higher than the limiting velocity of a FMDW and can be achieved at attainable currents densities. The NSOT induces a  force on a 90$^\circ$ AFDW  which is orders of magnitude higher compared to a Zeeman field and thus provides an effective control of the AF state. For a 180$^\circ$ rotated AF domains, only the staggered NSOT field can drive the AFDW.

We acknowledges support from Humboldt Foundation, EU ERC Advanced Grant No. 268066, from the Ministry of Education of the Czech Republic Grant No. LM2011026, and from the Grant Agency of the Czech Republic Grant no. 14-37427. We thank V. Amin for help with creating the graphics. 
%

\beginsupplement
\begin{widetext}
	\section*{Supplementary material}
	\section*{Equations for energy-momentum of an antiferromagnetic texture}
	In this Supplementary materials we show how to derive the dynamic equations for energy and momentum of an antiferromagnetic (AF) texture starting from the ideas of Refs.~\onlinecite{Haldane1983a, Kosevich1990, Ivanov1995}. We consider a collinear AF whose state is described by the N\'{e}el vector $\mathbf{L}(t,\mathbf{r})$ ($|\mathbf{L}|=M_s$). The orientation of the N\'{e}el vector can vary in space and time, hence we treat $\mathbf{L}(t,\mathbf{r})$ as a field variable.  For the sake of simplicity we assume that an AF sample is rather large and disregard boundary conditions. 
	
	\subsubsection{General considerations} The derivation is based on three ideas. First, an AF possess a nonzero magnetization which originates from the external magnetic field, $\mathbf{B}_\mathrm{Zee}$, and from the dynamics of the N\'{e}el vector. If the antiferromagnetic exchange coupling between the magnetic sublattices  (parametrized with the effective, so called spin-flip, field $H_\mathrm{ex}$) is much stronger than all other fields, magnetization of an AF can be explicitly expressed through the N\'{e}el vector as: \cite{Andreev1980,Baryakhtar1980}
	\begin{equation}\label{eq_magnetization}
	\mathbf{M}_\mathrm{AF}=\frac{1}{M_{s} H_\mathrm{ex}}\mathbf{L} \times \mathbf{B}_\mathrm{%
		Z e e} \times \mathbf{L}+\frac{\mathbf{L} \times\dot{\mathbf{L}}}{\gamma M_{s} H_\mathrm{ex}},
	\end{equation}
	where $\gamma$ is gyromagnetic ratio.
	Correspondingly, the equation of motion for the AF vector in the presence of spin pumping and damping can be treated as the balance equation for the magnetization $\mathbf{M}_\mathrm{AF}$:\cite{Baryakhtar1980, Gomonay2010,Gomonay2012}
	\begin{equation}\label{eq_dynamics_magnetization}
	\frac{d\mathbf{M}_\mathrm{AF}}{dt}=\gamma\mathbf{L}\times\mathbf{H}_\mathbf{L}-\gamma\alpha_G\mathbf{L}\times\dot{\mathbf{L}} +\boldsymbol{\Pi},
	\end{equation}
	where $\alpha_G$ is Gilbert damping constant, $\boldsymbol{\Pi}$ is a flux of magnetization which can originate e.g., from spin current transferred to the localized spins. The effective field
	\begin{equation}
	\mathbf{H}_\mathrm{L}\equiv-\frac{\partial w_\mathrm{AF}}{\partial \mathbf{L}}+\mathbf{B}_\mathrm{Neel}-\frac{ \mathbf{B}_\mathrm{Z e e}\left (\mathbf{L}
		\cdot \mathbf{B}_\mathrm{Zee}\right )}{H_\mathrm{ex}M_s}
	\end{equation}  
	is, in thermodynamic sense, conjugate to the N\'{e}el vector. It includes contributions from the N\'{e}el field $\mathbf{B}_\mathrm{Neel}$, Zeeman magnetic field $\mathbf{B}_\mathrm{Z e e}$, and magnetocrystalline anisotropy with the energy density $w_\mathrm{AF}$. 
	
	Second, if the external fields are relatively small (i.e. below the limit of the texture stability), a texture can move as a whole with only slight variation of its shape. At this level the texture can be treated as a particle and can be described with such variables as energy, momentum, angular momentum, which in the framework of classical mechanics are related to conserved quantities. \cite{Haldane1983a, Kosevich1990, Ivanov1995} 
	
	Third, dynamics of AF is invariant with respect to the Lorentz transformations, \cite{Haldane1983a, Kosevich1990, Ivanov1995} where the magnon velocity $c$ plays the role of limiting velocity of excitations in the media (equivalent to the speed of light). This can be seen immediately from the dynamic equation (deduced from (\ref{eq_magnetization}) and (\ref{eq_dynamics_magnetization}) ) for the infinite homogeneous AF in the absence of the external fields and damping:
	\begin{equation}\label{eq_motion_AF_ideal}
	\mathbf{L}\times\left[\ddot{\mathbf{L}}-c^2\Delta\mathbf{L} +\gamma^2H_\mathrm{ex}M_s\frac{\partial w_\mathrm{AF}}{\partial\mathbf{L}}\right]=0.
	\end{equation}
	Equation (\ref{eq_motion_AF_ideal}) has at least three integrals of motion (which coincide with the conservation laws): texture energy, $E$, momentum, $\mathbf{P}$, and angular momentum. We consider only two of them, energy 
	\begin{equation}\label{eq-first_integral_motion}
	E=\int \left[\frac{\dot{\mathbf{L}}^2+c^2(\nabla\mathbf{L})^2}{2\gamma^2M_sH_\mathrm{ex}}+w_\mathrm{AF}\right]dV,
	\end{equation} 
	and momentum 
	\begin{equation}\label{eq_canonical_momentum}
	P_j=-\frac{1}{\gamma^2M_sH_\mathrm{ex}}\int \dot{\mathbf{L}}\partial_j{\mathbf{L}}dV,\,j=x,y,z.
	\end{equation}
	It should be noted that due to the Lorentz invariance of Eq.~(\ref{eq_motion_AF_ideal}), $(E,\mathbf{P})$ can be treated as the components of a 4-vector. 
	
	In the presence of Zeeman and N\'{e}el fields the energy-momentum vector,  $(E,\mathbf{P})$, is no longer conserved, as 
	these fields produce the effective force 
	\begin{equation}
	\boldsymbol{\Gamma}=\gamma^2\mathbf{L}\times \left[H_\mathrm{ex}M_s\mathbf{B}_\mathrm{Neel}-\mathbf{B}_\mathrm{Zee}\left (\mathbf{L}
	\cdot \mathbf{B}_\mathrm{Zee}\right )\right]
	+\gamma\mathbf{L}\times\dot{\mathbf{B}}_\mathrm{Zee}\times\mathbf{L}-2\gamma \dot{\mathbf{L}}(\mathbf{B}_\mathrm{Zee}\mathbf{L}),
	\end{equation}
	which acts on the N\'{e}el vector. It should be noted that, as the dynamic equations for AFs are rather Newtonian-like than gyroscopic-like (like in FM), we can treat {$\boldsymbol{\Gamma}$} as a force, not as a torque.
	
	The balance equations for $(E,\mathbf{P})$ form a set of dynamic equations for an AF texture. To deduce these equations one starts from  general equation of motion for the N\'{e}el vector in the presence of the external fields and damping:
	\begin{equation}\label{eq_motion_AF}
	\mathbf{L}\times\left[\ddot{\mathbf{L}}-c^2\Delta\mathbf{L} +\gamma^2H_\mathrm{ex}M_s\frac{\partial w_\mathrm{AF}}{\partial\mathbf{L}}\right]
	=\boldsymbol{\Gamma}-\gamma\alpha_GH_\mathrm{ex}\mathbf{L}\times\dot{\mathbf{L}}.
	\end{equation}
	Once $\mathbf{L}(t,\mathbf{r})$ is the solution of the dynamic Eq.~(\ref{eq_motion_AF_ideal}), time derivatives of $E$ and $\mathbf{P}$ are calculated according to Eqs. (\ref{eq-first_integral_motion})  and (\ref{eq_canonical_momentum}) as follows:
	\begin{equation}\label{eq_energy_simplified_1}
	\frac{dE}{dt}=-\frac{\alpha_G}{\gamma M_s}\int\dot{\mathbf{L}}^2dV+\int\frac{\partial w}{\partial \mathbf{L}}\cdot\dot{\mathbf{L}}dV
	-\frac{1}{\gamma H_\mathrm{ex}}\int\dot{\mathbf{B}}_\mathrm{Z e e}\cdot\mathbf{L}\times\dot{\mathbf{L}}dV,
	\end{equation}
	and 
	\begin{equation}\label{eq_momentum_simplified_1}
	\frac{dP_j}{dt}=-\gamma\alpha_GH_\mathrm{ex}P_j-\int w(\mathbf{L})dS_j
	-		\frac{1}{\gamma M_sH_\mathrm{ex}}\int\dot{\mathbf{B}}_\mathrm{Z e e}\cdot\mathbf{L}\times\partial_j\mathbf{L}dV,
	\end{equation}
	where
	\begin{equation}\label{eq_field_energy_1}
	w(\mathbf{L})=-\frac{1}{2 M_{s} H_\mathrm{ex}}\left (\mathbf{L}
	\times \mathbf{B}_\mathrm{Z e e}\right )^{2}  -\mathbf{L} \cdot\mathbf{B}_\mathrm{%
		N e e l}
	\end{equation}
	is the energy density of the external fields, see Eq.~(1) of the main text. In Eqs. (\ref{eq_energy_simplified_1}) and (\ref{eq_momentum_simplified_1}) we have omitted the terms which vanish in the one dimensional  AF texture.
	The terms in the r.h.s. of Eqs. (\ref{eq_energy_simplified_1}) and (\ref{eq_momentum_simplified_1}) could be interpreted as follows. The first terms, proportional to $\alpha_G$, are associated with dissipation (deceleration) due to internal damping. The second terms in each equation, that depends on $w(\mathbf{L})$, stem from the pressure produced by the Zeeman and N\'{e}el fields. The last terms, proportional to $\dot{\mathbf{B}}_\mathrm{Zee}$, are related with  magnetization pumping induced by the time-dependent magnetic field. From  Eq.~(\ref{eq_momentum_simplified_1}) the  relaxation time of AF texture is defined as
	\begin{equation}\label{eq_relaxation time}
	\tau_{\rm{AF}}=\frac{1}{\alpha_G\gamma H_\mathrm{ex}}.
	\end{equation}
	
	\subsubsection{One dimensional texture} Any moving smooth AF texture can be viewed as a space/time rotation of an AF vector with respect to some reference configuration (e.g. its orientation at the sample boundary). In this cases it is convenient to parametrize the AF texture with the rotation angles and corresponding frequencies. In particular, space/time derivatives of the N\'{e}el vector can be expressed as
	\begin{equation}\label{eq_frequencies_parametrization}
	\dot{\mathbf{L}}=\boldsymbol{\Omega}_t\times\mathbf{L},\, \partial_j\mathbf{L}=\boldsymbol{\Omega}_j\times\mathbf{L},
	\end{equation}
	where the field variables $\boldsymbol{\Omega}_t(t,\mathbf{r})$ and $\boldsymbol{\Omega}_j(t,\mathbf{r})$ are time and space rotation frequencies. 
	
	\begin{figure}[h]
		\centering
		\includegraphics[width=0.7\linewidth]{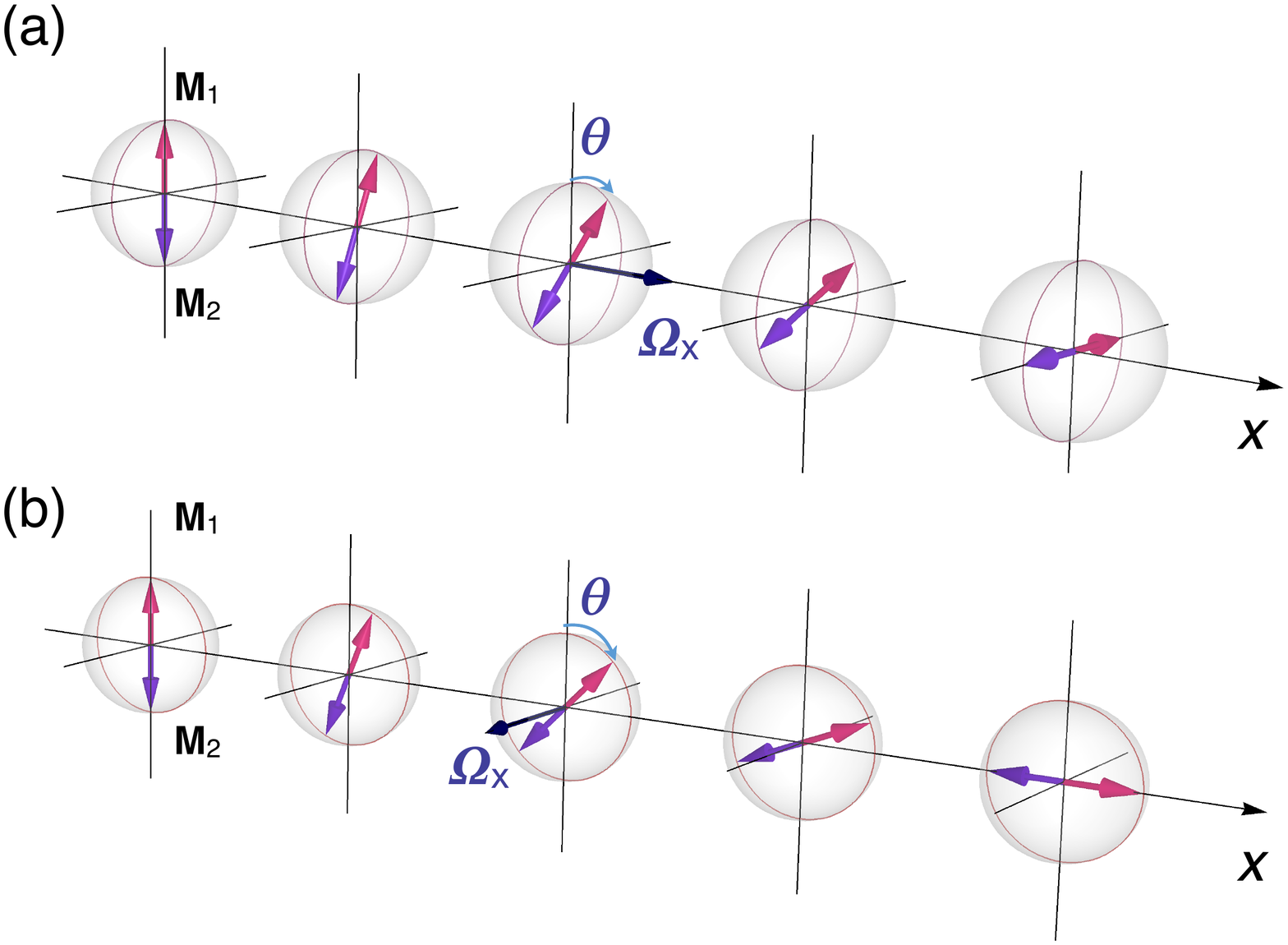}
		\caption{Bloch-like (a) and {N\'{e}el}-like (b) 90$^\circ$ domain walls in AF.  In both cases the AF vector {$\mathbf{L}=\mathbf{M}_1-\mathbf{M}_2$} varies along the $x$ axis. However, orientation of space rotational frequency $\boldsymbol{\Omega}_x$ is different: parallel to axis of inhomogenuity for the Bloch wall (a) and perpendicular to this axis for the N\'{e}el wall (b). }\label{fig_BLoch_vs_Neel_DW}
		\vskip -0.5 cm
	\end{figure}
	
	In the  1D  AF domain wall the N\'{e}el vector rotates in a fixed plane. In this simple case the vector $\boldsymbol{\Omega}_x$ (where $x$ is the direction of inhomogeneity) is oriented perpendicular to the rotation plane, and its value $|\boldsymbol{\Omega}_x|=\partial \theta/\partial x$ (see Fig.~\ref{fig_BLoch_vs_Neel_DW}). Suppose, the texture moves in the $x$ direction with the velocity $v$, keeping its shape almost constant, i.e., $\mathbf{L}(\mathbf{r}-\mathbf{v}t)$ satisfies, at least approximately, the equation of motion (\ref{eq_motion_AF_ideal}). In this case $\boldsymbol{\Omega}_t=-v\boldsymbol{\Omega}_x/\sqrt{1-v^2/c^2}$, where we took into account relativistic character of AF dynamics. Then, from Eq.~(\ref{eq-first_integral_motion}) we get 
	\begin{equation}\label{eq_energy_relativistic}
	E=S\int \left[\frac{M_s}{2\gamma^2H_\mathrm{ex}}\frac{c^2}{1-v^2/c^2}\left(\frac{\partial\theta}{\partial x}\right)^2+w_\mathrm{AF}\right]dx,
	\end{equation} 
	where $S$ is the square of AF sample in the $yz$ plane. 
	As $\mathbf{L}(\mathbf{r}-\mathbf{v}t)$ satisfies equation of motion (\ref{eq_motion_AF_ideal}), the conditions of the virial theorem are fulfilled (the averaged kinetic energy is equal to the half of the averaged potential energy of the wall), and then,
	\begin{equation}\label{eq_energy_relativistic_1}
	E=\frac{M_sS}{\gamma^2H_\mathrm{ex}}\frac{c^2}{1-v^2/c^2}\int_{-\infty}^{\infty} \left(\frac{\partial\theta}{\partial x}\right)^2dx.
	\end{equation}
	The corresponding momentum is
	\begin{equation}\label{eq_momentum_relativistic}
	P_x=\frac{v}{\sqrt{1-v^2/c^2}}\frac{M_sS}{\gamma^2H_\mathrm{ex}}\int_{-\infty}^{\infty} \left(\frac{\partial\theta}{\partial x}\right)^2dx.
	\end{equation}
	So, the value
	\begin{equation}\label{eq_mass}
	M_\mathrm{AFDW}=\frac{M_sS}{\gamma^2H_\mathrm{ex}}\int_{-\infty}^{\infty} \left(\frac{\partial\theta}{\partial x}\right)^2dx
	\end{equation}
	can be interpreted as an effective mass of the AF texture. The value of the integral in Eqs. (\ref{eq_energy_relativistic_1})-(\ref{eq_mass}) depends upon particular magnetic symmetry and domain wall type. In the simplest case of 180$^\circ$ domain wall with $\theta(t,x)$ satisfying Eq.~(2) of the main text, $\int(\partial \theta/\partial x)^2dx=1/x_\mathrm{DW}$ and $M_\mathrm{AFDW}=M_sS/(\gamma^2H_\mathrm{ex}x_\mathrm{DW})$.
	
	The dynamic equations for energy (\ref{eq_energy_simplified_1}) and momentum (\ref{eq_momentum_simplified_1}) with account of the relations (\ref{eq_energy_relativistic_1}) and  (\ref{eq_momentum_relativistic}) then take a final form:
	\begin{equation}\label{eq_energy_simplified_2}
	\frac{dE}{dt}=-\frac{v^2}{c^2}\gamma\alpha_GH_\mathrm{ex}E+\frac{vS}{\sqrt{1-v^2/c^2}}\left[w(\mathbf{L}_1)-w(\mathbf{L}_2)\right]
	-\frac{vM_sS}{\gamma H_\mathrm{ex}\sqrt{1-v^2/c^2}}\int_{-\infty}^{\infty} \dot{\mathbf{B}}_\mathrm{Zee}\cdot\boldsymbol{\Omega}_xdx,
	\end{equation} 
	and
	\begin{equation}\label{eq_momentum_simplified_2}
	\frac{dP_x}{dt}=-\gamma\alpha_GH_\mathrm{ex}P_x+S\left[w(\mathbf{L}_1)-w(\mathbf{L}_2)\right]
	-\frac{M^2_sS}{\gamma H_\mathrm{ex}}\int_{-\infty}^{\infty} \dot{\mathbf{B}}_\mathrm{Zee}\cdot\boldsymbol{\Omega}_x dx,
	\end{equation}
	where the vectors $\mathbf{L}_{1,2}\equiv \mathbf{L}(t=0,x=\mp \infty)$.

	The parameters of steady motion (velocity, momentum and energy) are calculated according to Eqs.~ (\ref{eq_energy_simplified_2}) and (\ref{eq_momentum_simplified_2}) assuming that $dE/dt=0$, $dP_x/dt=0$. 
	
	Equation (\ref{eq_momentum_simplified_2})  coincides with the Eq.~(8) of the main text, the effective force per unit area ($S=1$) being
	\begin{equation}\label{eq_effective force}
	F_x=w(\mathbf{L}_1)-w(\mathbf{L}_2)
	-\frac{M^2_s}{\gamma H_\mathrm{ex}}\int \dot{\mathbf{B}}_\mathrm{Zee}\cdot\boldsymbol{\Omega}_x dx.
	\end{equation} 
	
	It should be noted that the equation for the collective coordinates proposed in Ref.~\onlinecite{Tveten2013} can be deduced from Eq.~(\ref{eq_momentum_simplified_2}) in the nonrelativistic limit ($v\ll c$).
	
	In the case of static field the second term in Eq.~(\ref{eq_effective force}) vanishes and 
	\begin{equation}\label{eq_effective force_static}
	F_x=\frac{\left (\mathbf{L}_1
		\mathbf{B}_\mathrm{Z e e}\right )^{2}-\left (\mathbf{L}_2
		\mathbf{B}_\mathrm{Z e e}\right )^{2}}{2 M_{s} H_\mathrm{ex}} +\left(\mathbf{L}_2-\mathbf{L}_1\right) \cdot\mathbf{B}_\mathrm{%
		N e e l}.
	\end{equation}
	Equation (9) of the main text for the velocity of steady motion is then obtained from Eqs.~(\ref{eq_momentum_relativistic}) and (\ref{eq_momentum_simplified_2}).   
	
	\subsubsection{Effective mass and relaxation time in AF and FM} While both the mass and relaxation time are strongly suppressed in the AF due to exchange, the ratio
	\begin{equation}
	\label{eq_mass_time_ratio_AF}
	\frac{M_\mathrm{AFDW}}{\tau_\mathrm{AF}}=\frac{\alpha_GM_sS}{\gamma x_\mathrm{DW}}
	\end{equation}
	is independent of $H_\mathrm{ex}$, as can be seen from Eqs.~(\ref{eq_relaxation time}) and (\ref{eq_mass}). In this subsection we show that  in the  FMDW the mass-to-time ratio is the same as in AFDW.
	
	The expression for the effective mass of the FMDW was introduced in Ref.~\onlinecite{Schlomann1972}. When written in the notation of the present paper  it takes the form (assuming that anisotropy and  dipolar anisotropy fields are comparable):
	\begin{equation}
	\label{eq-FM_mass}
	M_\mathrm{FMDW}=\frac{M_sS}{\gamma^2H_\mathrm{ex}x_\mathrm{DW}}.
	\end{equation}
	The  relaxation time of a FM can be derived from the definition of Gilbert damping constant as an inverse quality factor of the FM resonance:
	\begin{equation}\label{eq_FM_time}
	\tau_\mathrm{FM}=\frac{1}{\alpha_G\omega_\mathrm{FM}}=\frac{1}{\gamma\alpha_G H_\mathrm{an}.}
	\end{equation}
	From Eqs.~(\ref{eq-FM_mass}) and (\ref{eq_FM_time}) one gets
	\begin{equation}
	\label{eq_mass_time_ratio_FM}
	\frac{M_\mathrm{FMDW}}{\tau_\mathrm{FM}}=\frac{\alpha_GM_sS}{\gamma x_\mathrm{DW}},
	\end{equation}
	which coincides with the expression (\ref{eq_mass_time_ratio_AF}).

\end{widetext}

\end{document}